\documentclass{emulateapj}
\RequirePackage{natbib}

\def\agt{\mathrel{\raise.3ex\hbox{$>$}\mkern-14mu\lower0.6ex\hbox{$\sim$}}}
\def\alt{\mathrel{\raise.3ex\hbox{$<$}\mkern-14mu\lower0.6ex\hbox{$\sim$}}}

\newcommand{\beq}{\begin{equation}}
\newcommand{\eeq}{\end{equation}}
\newcommand{\beqn}{\begin{eqnarray}}
\newcommand{\eeqn}{\end{eqnarray}}

\shorttitle{Bar-Mode Instability}
\shortauthors{Shapiro}

\begin{document}

\title{The Secular Bar-Mode Instability 
in Rapidly Rotating Stars Revisited}

\author{Stuart L. Shapiro \altaffilmark{1,2}}

\affil
{\altaffilmark{1} 
Department of Physics, University of Illinois at Urbana-Champaign,
\break
Urbana, IL 61801-3080}
\affil
{\altaffilmark{2} 
Department of Astronomy and NCSA, University of Illinois at
Urbana-Champaign,
\break
Urbana, IL 61801-3080}

\begin{abstract}
Uniformly rotating, homogeneous, incompressible Maclaurin spheroids
that spin sufficiently rapidly are secularly unstable to nonaxisymmetric, 
bar-mode perturbations when viscosity is present. The intuitive explanation is
that energy dissipation by viscosity can drive an unstable spheroid to 
a stable, triaxial configuration of lower energy -- a Jacobi ellipsoid.
But what about rapidly rotating {\it compressible} stars? Unlike incompressible
stars, which contain no internal energy and therefore immediately liberate 
all the energy dissipated by viscosity, compressible stars have internal energy and can 
retain the dissipated energy as internal heat. 
Now compressible stars that rotate sufficiently rapidly {\it and} also manage to liberate this 
dissipated energy very quickly are known to be unstable to bar-mode perturbations, like 
their incompressible counterparts. But what is the situation for rapidly rotating compressible stars
that have very long cooling timescales, so that all the energy dissipated by viscosity 
is retained as heat, so that the total energy of the star remains constant on a 
secular (viscous) evolution timescale? 
Are such stars also unstable to the nonlinear growth of bar modes, or is the viscous heating 
sufficient to cause them to expand, drive down the ratio of
rotational kinetic to gravitational potential energy $\mathcal {T}/|W| \propto R_{\rm eq}^{-1}$, 
where $R_{\rm eq}$ is 
the equatorial radius, and turn off the instability before it gets underway? Alternatively, 
if the instability still arises in such stars, at what rotation rate do they
become unstable, and to what final state do they evolve?  We provide definitive answers to these
questions in the context of the compressible ellipsoid model for rotating stars.
The results should serve as useful guides for
numerical simulations that solve the exact Navier-Stokes equations in $3+1$ dimensions for 
rotating stars containing viscosity.
\end{abstract}




\section{Introduction}

Rapidly spinning equilibrium stars are subject to nonaxisymmetric
rotational instabilities.  An exact treatment of these instabilities
exists for incompressible fluid configurations in Newtonian
gravity (see, e.g., Chandrasekhar 1969). 
For axisymmetric, uniformly rotating Maclaurin spheroids, global
rotational instabilities arise from nonradial toroidal modes
$e^{im\varphi}$ ($m=\pm 1,\pm 2, \dots$) when $\mathcal{T}/|W|$ exceeds a
certain critical value. Here $\varphi$ is the azimuthal coordinate and
$\mathcal{T}$ and $W$ are the rotational kinetic and gravitational potential 
energies of the star, respectively.  
In the following discussion we will focus on the $m=\pm 2$ bar-mode,
also known as the $l=m=2$ $f$-mode ($f=$ fundamental),
which is the fastest growing mode when the rotation is sufficiently
rapid.

There exist two different mechanisms and corresponding timescales for
bar-mode instabilities.  Uniformly rotating, incompressible stars in
Newtonian theory are {\it secularly} unstable to bar-mode formation
when $\mathcal{T}/|W| = 0.1375$.  This instability can
grow only in the presence of some dissipative mechanism, like
viscosity or gravitational radiation, and the growth time occurs on
the dissipative timescale, which is usually much longer
than the dynamical timescale of the system.  By contrast, a {\it
dynamical} instability to bar-mode formation sets in when 
$\mathcal{T}/|W| = 0.2738$.  This instability is independent of any
dissipative mechanisms, and the growth time occurs on the
hydrodynamical timescale of the system. Here we shall be interested in
the {\it secular} bar-mode driven by \mbox{\it viscous} dissipation.

Nonaxisymmetric secular instabilities in uniformly rotating compressible stars, 
e.g., polytropes, have been analyzed by several authors 
(see, e.g., Managan 1985; Imamura, Friedman \&
Durisen 1985; Ipser \& Lindblom 1990, 1991). The $m=2$ bar-mode sets in again at $\mathcal{T}/|W| \simeq 0.14$. However, this mode
is reached only when the polytropic index of the star satisfies 
$n \lesssim 0.808$ (James 1964). Stars with larger  $n$ are too 
centrally condensed to support high enough
spin in uniform rotation without undergoing mass-shedding at the
equator (see Tassoul 1978 and Shapiro \& Teukolsky 1983 
for a discussion and references). This constraint does not apply to
differentially rotating stars, which can support significantly
more rotational energy in equilibrium, even when the degree of
differential rotation is only moderate.
The critical value for the onset of the 
secular $m=2$ bar-mode is again $\mathcal{T}/|W| \simeq 0.14$ for a wide range of angular
momentum distributions and barotropic equations of state
(see, e.g., Ostriker \& Bodenheimer 1973; Bardeen et al. 1977; 
Friedman \& Shutz 1978a,b). Hence stability criteria derived 
for uniformly rotating, incompressible spheroids turn out to 
have general applicability to rapidly rotating stars with more 
realistic density and angular velocity profiles, at least insofar as
the $m=2$ bar-mode is concerned.

The point of onset of the secular bar-mode instability in homogeneous, 
axisymmetric Maclaurin spheroids exactly coincides with the 
point of bifurcation where the Maclaurin
sequence branches off into the nonaxisymmetric Jacobi sequence. 
Incompressible Jacobi ellipsoids are homogeneous, uniformly rotating, triaxial equilibria that have 
lower total energies than corresponding Maclaurin spheroids of the same
mass, density and angular momentum.
The nonlinear,
secular evolution of a slightly perturbed, unstable Maclaurin spheroid to
a triaxial Jacobi ellipsoid was explicitly demonstrated by
Press and Teukolsky (1973), who integrated
the full set of Riemann-Lebovitz ordinary differential
equations for incompressible ellipsoids with viscosity.
Sequences of uniformly rotating polytropes with $n \leq 0.808$ 
also have a point of bifurcation beyond which they are secularly unstable;
for example, this point occurs at $\mathcal{T}/|W| = 0.1298$ for
$n=0.75$ (Ipser \& Lindblom 1990), close to the
value $\mathcal{T}/|W| = 0.1375$ for $n = 0$ (i.e. incompressible) stars. Viscosity 
again is expected to drive unstable models to secularly stable,
Jacobi-like ellipsoids in compressible stars.

It seeems almost self-evident why spheroids beyond the bifurcation point 
are secularly unstable to 
bar-mode perturbations, since with viscosity such configurations can evolve 
to equilibrium configurations of lower energy, while 
conserving their mass and angular momentum. However, this intuitive explanation
is not applicable to a rapidly rotating star
in which the energy liberated by viscosity
is {\it not} emitted by the star, but instead remains trapped  
in the form of thermal energy. In this case
the star loses no mass, angular momentum {\it or} energy as it evolves via
viscosity,
and thus it cannot evolve to a lower-energy equilibrium state. 
This problem does not arise in
incompressible stars, since they possess no internal (thermal) energy 
whatsoever and, hence, all the energy generated by viscosity is liberated
instantaneously.  The rate of decrease in the total energy of an 
incompressible star is exactly the rate of viscous energy generation;
this equality is guaranteed by the Riemann-Lebovitz equations 
with viscosity.  By contrast, compressible matter can possess thermal 
energy, and the fate of
viscous energy dissipation in compressible stars 
depends on timescales. If the cooling 
timescale, due, e.g., to thermal radiation, is shorter than the 
thermal heating timescale due to viscosity, then the thermal energy
generated by viscosity  will be radiated away quickly 
and the total energy of the star
will decrease on the viscous dissipation timescale, as in the incompressible
case. If, however, the cooling timescale is longer than the viscous 
heating timescale, 
there will be an increase in thermal energy, but the total energy of the
star will be nearly constant on a viscous timescale. 
The possibility for the nonlinear growth of a bar-mode instability is not so 
obvious in the later situation.
In fact, in a compressible star in which the cooling timescale 
is long, any viscosity-generated enhancement in the 
thermal energy might cause
the star to expand, increasing the equatorial radius $R_{\rm eq}$, 
decreasing the value of $\mathcal{T}/|W| \sim 1/R_{\rm eq}$ and potentially 
turning off any nonaxisymmetric mode that might be unstable initially. 

In this paper we address the issue of the secular bar-mode instability 
in rapidly rotating compressible stars in which the cooling timescale is
either much longer, or much shorter, than the viscous heating timescale. 
We integrate the dynamical equations for a compressible star with viscosity
using the formalism of Lai, Rasio \& Shapiro (1994) (hereafter LRS 1994). 
This formalism
provides a set of {\it ordinary} differential equations (ODEs) for the evolution of the
principal axes and other global parameters characterizing the star, 
which is approximated as a triaxial, compressible ellipsoid with a 
polytropic equation of state and a velocity field that is a linear function 
of the coordinates. The 
equations are exact in the incompressible limit $n=0$, where
they describe the exact hydrodynamical behavior of the objects and 
reduce to Riemann-Lebovitz equations identically.
This dynamical model for compressible
ellipsoids is formally equivalent to the affine
model developed by Carter \& Luminet (1983,1985), although 
the formulation is quite different. 

We follow the evolution of rapidly rotating, 
compressible spheroids that lie beyond the bifurcation point 
and are given small,
nonaxisymmetric perturbations at $t=0$. 
We track their evolution in the two extreme opposite limits of
very rapid and very slow cooling. We find that in {\it both} cases the 
configurations are secularly unstable to a bar-mode perturbation, but 
they evolve to {\it different}, stable, compressible Jacobi ellipsoids. 
In the rapid-cooling limit, the final ellipsoid has lower energy than the 
initial spheroid, while in the slow-cooling limit, the final 
ellipsoid  has the same energy, but a higher entropy than the
spheroid.

Our paper is organized as follows. In Section 2 we describe the compressible
ellipsoid model for rotating stars and extend the formalism of LRS to the
case in which the entropy of the fluid is allowed to change with time due to heating and cooling.
In Section 3 we integrate the dynamical equations with viscosity to show that, both in the case
of very rapid and very slow cooling, a uniformly rotating, compressible spheroid that rotates sufficiently
rapidly and is subjected to a small triaxial perturbation 
is unstable to the bar instability and evolves to a Jacobi ellipsoid. In this section we
also show that, in fact, the final Jacobi ellipsoid, which is different in the two cases, can be determined
a priori from the conserved quantities without having to follow the evolutionary track.
In Section 4 we confirm our numerical result that, in both cases, 
the onset of secular instability coincides with
the bifurcation point. In particular, we sketch a proof of this result by examining
the energy and free energy functionals of perturbed stars along the compressible Maclaurin sequence.
Finally, in Section 5 we summarize our results and describe plans for future work.

\section{Basic Equations}
\label{Sec2}

The key dynamical equations for the evolution of a compressible
fluid in the ellipsoidal approximation
are derived in LRS (1994)
and will not be repeated here. [See Section 2 of that paper for 
an ideal fluid, and Section 4 for a fluid with shear
viscosity.] They are obtained from the Euler-Lagrange equations derived
from a Lagrangian governing the dynamics of a compressible ellipsoid. 
The properties of compressible equilibria satisfying the
stationary-state ellipsoidal equations are described in LRS (1993). The 
assumptions underlying the compressible ellipsoidal treatment, as well as comparisons between
ellipsoidal model calculations and calculations employing the exact fluid
equations, are summarized in these papers and the references therein. The basic assumption 
is that we assume that the surfaces of constant density can be represented approximately
by {\it self-similar ellipsoids}. The geometry is then completely specified by the three principal
axes of the outer surface. Furthermore, we assume that the density profile $\rho(m)$ inside each star,
where $m$ is the mass interior to an isodensity surface, is identical to that of a {\it spherical}
polytrope with the same volume. The velocity field of the fluid in a frame at rest with the star's 
center of mass is modeled as a linear superposition
of three components: (1) a rigid rotation of the ellipsoidal {\it figure}; (2) an internal fluid circulation
with {\it uniform vorticity}; and (3) ellipsoidal expansion or contraction. In our adaptation, where we restrict
ourselves to the cases in which both the angular velocity and the vorticity are parallel to one of the principal
axes, the exact
hydrodynamical equations (the Euler and Navier-Stokes equations) are replaced by a set of ODEs for the
time evolution of the principal axes, the angular velocity of the ellipsoidal figure and the angular frequency
of the internal circulation. As mentioned above, in the incompressible limit (polytropic index $n=0$) the solutions
we derive represent {\it exact} solutions of the true hydrodynamic equations. For $n \neq 0$, our solutions
are only approximate, since the isodensity surfaces can no longer be exactly ellipsoidal, and the velocity field
of the fluid cannot be described exactly by a linear function of the coordinates. We adopt
the notation of LRS in the discussion below.

\subsection{Energy Evolution Equation}
\label{Secevo}

The equation of state characterizing the fluid is given by the 
polytropic law
\begin{equation} \label{eos}
P = K \rho^{\Gamma}, ~~~\Gamma = 1 + 1/n,
\end{equation}
where $P$ is the pressure, $\rho$ is the density, $\Gamma$ is the adiabatic
constant and $n$ is the polytropic index.
If the polytropic entropy parameter $K$ is assumed to be
constant in space {\it and} time, as in LRS (1994), the equation of state is 
barotropic with $P = P(\rho)$. In this case, no internal energy
equation is needed to describe the evolution of the 
thermodynamic state of the star.
In general, however, the gas may undergo heating and cooling, so
that  the parameter $K$ will be a function of the changing specific
entropy $s$, i.e., $K = K(s)$.  In this case the equation of state
is not barotropic, since $P = P(\rho, s)$, and it is necessary to solve an
energy evolution equation, which we will now derive. Combining 
eqn.~(\ref{eos}) with
the first law of thermodynamics,
\begin{equation} \label{first}
T ds = d (\varepsilon/\rho) + P d(1/\rho),
\end{equation}
where $T$ is the temperature, and adopting the 
ideal gas relation for the internal energy density, 
$\varepsilon = P/(\Gamma -1)$,  
yields an equation for $K$:
\begin{equation} \label{K}
T ds = \frac{\rho^{\Gamma-1}}{\Gamma-1} dK.
\end{equation}
Henceforth we shall assume, for simplicity,
that $s$, and hence $K$, are independent of position
in the star, but we will allow them to vary with time.
Integrating eqn.~(\ref{K}) over the entire compressible ellipsoid, 
yields an evolution equation
for $K$:
\begin{equation} \label{kdot}
U \frac{d \ln K}{dt} = {\langle T \rangle} \frac{dS}{dt} 
= \Gamma_{\rm vis} - \Lambda_{\rm cool}.
\end{equation}
In eqn.~(\ref{kdot}), $\langle T \rangle$ is the mass-averaged (mean) temperature defined by 
\begin{equation} \label{temp}
\langle T \rangle = \frac{1}{M} \int T dm,
\end{equation}
$S = sM$ is the total entropy, $M$ is the total mass, and $U$ is
the total internal energy of the star, given by 
\begin{equation} \label{u}
U = \int \left ( \varepsilon/ \rho \right) dm = k_1 K \rho_c^{1/n} M , 
\end{equation}
(eqn. 2.9 in LRS 1994), where
$k_1$ is a dimensionless polytropic structure constant of order unity 
which depends on the polytropic
index $n$ ($ k_1 \sim n$) and $\rho_c$ is the central density.
As a simple example, for the case of an isentropic
ellipsoid described by a Maxwell-Boltzmann equation of state, we have
\begin{eqnarray} \label{maxb}
P &=& \rho k_B T/\mu, ~~~~ K = \rho_c^{1-\Gamma} k_B T_c/\mu, \nonumber \\
U &=& k_1 k_B T_c M/\mu,  ~~~~ \langle T \rangle = k_1 (\Gamma-1) T_c, \\
S &=& U \ln K /\langle T \rangle = \frac{ k_B}{\Gamma - 1} \nonumber
\frac{M}{\mu} \ln K.
\end{eqnarray}
In eqn.~(\ref{maxb}) $k_B$ is Boltzmann's constant, $\mu$ is the mean
molecular weight and the subscript `c'
denotes that the quantity is evaluated at the center of the star.

The quantity  $\langle T \rangle dS/dt$ in eqn.~(\ref{kdot}) 
is the {\it net} heating
rate of the star, determined by the difference between the 
heating rate due to viscosity, $\Gamma_{\rm vis}$, and 
the cooling rate, $\Lambda_{\rm cool}$,
due to radiation losses. The viscous heating rate is given by
\begin{equation} \label{heat}
\Gamma_{\rm vis} = \int \sigma_{ij}u_{i,j} d^3 x = -{\cal W},
\end{equation}
where $u_i$ are the components of the stellar velocity,
$\sigma_{ij}$ is the viscous stress tensor and
${\cal W}$ is the viscous dissipation rate. In eqn.~(4.6) of 
LRS (1994) the quantity
${\cal W}$ is calculated analytically in terms of the principal axes of the
ellipsoid $a_i$, their velocities $\dot a_i$ and the angular frequency
of the internal fluid motion $\Lambda$ ($\propto$ vorticity), and is proportional 
to the mass-averaged
shear viscosity ${\langle \nu \rangle}$ defined by
\begin{equation} \label{vis}
\langle \nu \rangle = \frac{1}{M} \int \nu dm .
\end{equation}
Following Press and Teukolsky (1973) we parametrize $\langle \nu \rangle$ according
to $\langle \nu \rangle = C_{\rm visc}(\pi G \bar \rho)^{1/2}R^2$, where $R = (a_1 a_2 a_3)^{1/3}$
is the mean radius of the ellipsoid, $\bar \rho = M/(4\pi R^3/3)$ is the mean
density and $C_{\rm visc}$ is a dimensionless constant.
Choosing $C_{\rm visc} \lesssim 1$ guarantees that the viscous damping timescale
will be longer than the hydrodynamical timescale of the star.

The form of the cooling rate $\Lambda_{\rm cool}$ will depend on the
details of the stellar microphysics, but here we shall only assume that
$\Lambda_{\rm cool} = 0$ when $K$ is held at its initial value $K(s) = K_0 = K(s_0)$ in
the absence of viscosity.
Thus, we assume that the fluid would evolve adiabatically on the timescales of interest
were the viscosity to be turned off completely.
For example, if we take the initial configuration to be a cold, degenerate neutron star
then we would have $K=K_0 > 0$ and no cooling or heating in the absence of viscosity
(i.e. no emission from a zero-entropy gas).

Setting $K = K_0$ at $t=0$, we shall  
treat the subsequent energy evolution in two extreme, opposite
physical regimes. In the ``rapid-cooling'' regime the cooling rate
greatly exceeds the viscous heating rate for all $K > K_0$. In this case
any thermal energy generated by viscosity will be radiated almost 
immediately, and the gas will not be heated much above $K_0$. 
In the limit of arbitrarily fast cooling,  eqn.~(\ref{kdot}) establishes
a balance between heating and cooling, i.e.,
$\Gamma_{\rm vis} = \Lambda_{\rm cool}$, and maintains the value of
$K$ arbitrarily close to $K_0$. We thus have 
\begin{equation} \label{rcool}
 \Gamma_{\rm vis} = \Lambda_{\rm cool}, \ \ \ \  K = K_0 ~{\rm = \text{constant}},\ \ \ \ {\rm (\text{``rapid \ cooling''})}.
\end{equation}
In the ``no-cooling'' regime, the viscous heating rate
greatly exceeds the cooling rate, in which case cooling is unimportant and
eqn.~(\ref{kdot}) gives
\begin{equation} \label{ncool}
\frac{d \ln K}{d t } = \Gamma_{\rm vis}/U = -{\cal W}/U > 0, 
~~~~~~{\rm (\small{``no \ cooling''})}.
\end{equation}
In this case the energy generated by viscosity will go into thermal
energy, which will be reflected in the secular increase of $K$ above its
initial value $K_0$.

Combining the general energy evolution eqn~(\ref{kdot}) with the
dynamical equations for a compressible ellipsoid 
(LRS 1994, eqns. 4.10- 4.14) yields a conserved energy integral:
\begin{equation} \label{cons}
E(t) + \int_{0}^{t} \Lambda_{\rm cool} dt 
= \ E(0) =\ {\rm constant},
\end{equation}
where the total energy $E(t)$ is given by
\begin{equation} \label{E}
E(t) = U + W + \mathcal{T}.
\end{equation}
In the ``rapid-cooling'' regime the total energy $E(t)$ of the star will
decrease secularly by exactly the amount of energy dissipated by
viscosity, according to eqns.~(\ref{rcool}) and ~(\ref{cons}).
This is {\it always} the situation in 
incompressible stars, since they
contain no internal energy and hence must lose energy whenever there is
viscous dissipation. In the ``no-cooling'' regime, by contrast, 
the energy $E(t)$ will be
strictly conserved, as the differential rotational
kinetic energy dissipated by viscosity is entirely converted into internal
energy and is not lost to the star. Compressible stars can
evolve in either of these extreme opposite regimes, as well as in 
intermediate regimes, depending on the physical conditions.
In all cases, the mass and angular momentum of a star is conserved in
the presence of viscosity, but circulation is not. These quantities are
all monitored in our evolution calculations.

In reporting the results of our calculations below, it will be
useful to define the following nondimensional
ratios for energy and angular momentum: 
\begin{equation} \label{dim}
\bar E = \frac{E}{G M^2/R_0}, ~~~\bar J = \frac{J}{(G M^3 R_0)^{1/2}},
\end{equation}
where
\begin{eqnarray} \label{rad}
R_0 & = & \xi_1 \left(\xi_1^2 |\theta'_1|\right)^{-(1-n)/(3-n)} 
\left(\frac{M}{4 \pi}\right)^{(1-n)/(3-n)} \nonumber \\
& & \times \ \left[\frac{(n+1) K_0}{4 \pi G}\right]^{n/(3-n)}
\end{eqnarray}
is the radius of a nonrotating {\it spherical} polytrope
of mass $M$, polytropic entropy parameter $K_0$ and polytropic index $n$.
The dimensionless quantities $\theta_1$ and $\xi_1$ are the familiar
Lane-Emden variables at the surface of a polytrope
(see, e.g. Chandrasekhar 1939).

\section{Secular Evolution of Unstable Spheroids}

We now determine the secular evolution of compressible ellipsoids with
viscosity in the ``rapid-cooling'' and ``no-cooling'' regimes.
Our initial configurations consist of compressible Maclaurin
spheroids that are given small, but finite, 
nonaxisymmetric (triaxial) perturbations and
evolved until they settle down to a final equilibrium state.
Typical results for secularly unstable (but dynamically stable) spheroids 
are shown in Fig. 1. The parameters $\{ a_i \}$ are the semi-major axes of the configuration, with
$a_3$ along the rotation axis; in axisymmetry $a_1 = a_2$.
For the models plotted in this figure we took
an $n=1$ compressible Maclaurin spheroid 
with eccentricity $e = (1 - a_3^2/a_1^2)^{1/2} = 0.94$ 
and $\mathcal{T}/|W| = 0.252$. We increased the equatorial axis $a_1$ slightly 
to induce a perturbation, setting $a_2/a_1 = 0.999$ at the 
start of the evolution. The viscosity parameter was chosen to be
$C_{\rm vis}=0.15$;  however, as long as the secular timescale is longer than
the dynamical timescale (i.e. as long as $C_{\rm vis}$ is sufficiently small), 
the secular (viscous) evolution track is independent of $C_{\rm vis}$ 
provided we ``rescale'' the evolution time according to 
$t \propto 1/C_{\rm vis}$. For both the ``rapid-cooling'' and
``no-cooling'' cases we find that the evolution proceeds quasi-statically 
through a sequence of compressible, differentially rotating, Riemann-S equilibrium ellipsoids, finally
arriving at a uniformly rotating, compressible, Jacobi ellipsoid. While the initial spheroid
is unstable in either cooling regime, 
both the nonlinear evolutionary track and the final Jacobi equilibrium 
ellipsoids are different in the two cases. 

\begin{figure}
\plotone{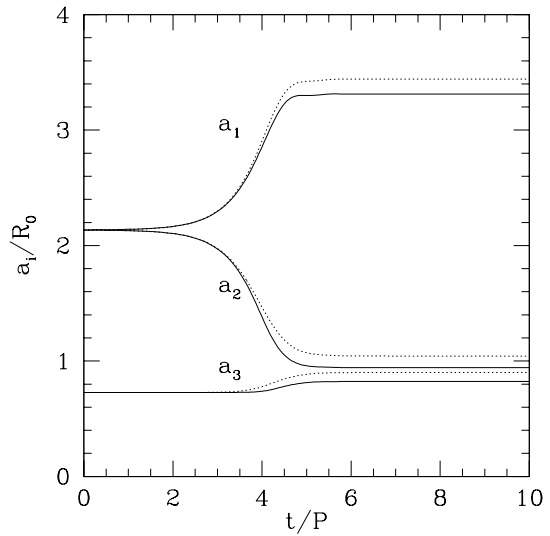}
\caption{Viscous evolution of a secularly unstable, uniformly rotating Maclaurin spheroid
with $e=0.94$ to a stable, uniformly rotating, Jacobi ellipsoid. The configuration is a compressible polytrope of
index $n=1$ that is given a slight nonaxisymmetric perturbation at $t=0$.  The solid lines correspond to
``rapid cooling'' and the dotted lines to ``no cooling''. Time is measured in units of the initial rotation period
of the spheroid, $P$, and the semi-major axes $a_i$ are measured in units of $R_0$ defined by eqn.~(\ref{rad}).
\label{fig1}}
\end{figure}

The difference in evolution in the two cases 
can be appreciated by examining Fig. 2, which plots 
isentropic (i.e. constant $K$) 
equilibrium curves for two sequences of compressible, $n=1$, Maclaurin spheroids 
of different entropy, and the corresponding Jacobi ellipsoids that branch off from these sequences. 
The initial spheroid for the evolution
plotted in Fig. 1 is indicated by the
solid dot on the $K=K_0$ Maclaurin curve beyond the 
bifurcation point.  For the case with ``rapid cooling'', 
eqn.~(\ref{rcool}) applies; accordingly, $M$, $J$ and $K$ all remain constant 
with time while the
total energy of the configuration decreases, as the energy dissipated by
viscosity escapes. The star thus evolves vertically downward in the
figure to a Jacobi ellipsoid with lower energy $E$ but
with the same angular momentum $J$ and entropy parameter $K=K_0$ as the initial configuration (solid triangle). 
By contrast, for the case with``no cooling'', eqn.~(\ref{ncool})
applies; thus, while $M$ and $J$ are again conserved, 
the entropy parameter $K$ increases with time, since the 
energy dissipated by viscosity is retained and heats up the star.
Consequently, the star evolves at constant $J$ {\it and} $E$ 
and does not move at all from its initial position in Fig. 2. However,
the configuration does change -- it evolves to the Jacobi ellipsoid which bifurcates from a
Maclaurin sequence with a higher entropy.  
The final Jacobi ellipsoid in this case has the
same $E$ and $J$ as the initial Maclaurin spheroid, but has a higher entropy parameter
$K > K_0$. For the example illustrated in Fig 1., $K/K_0 = 1.197 > 1$
for this final ellipsoid. We note that unstable {\it incompressible} spheroids
always evolve in the ``rapid-cooling'' regime, since they have no
internal energy and cannot retain viscous dissipation as heat. 

\begin{figure}
\plotone{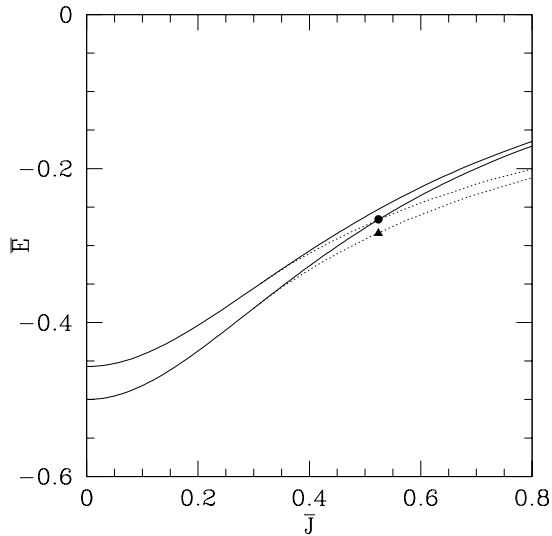}
\caption{Energy as a function of angular momentum for two isentropic, equilibrium sequences of compressible spheroids
(solid lines) and the compressible Jacobi ellipsoids which branch off from them (dotted lines). The 
lower sequences have an entropy constant $K_0$, while the upper sequences have $K = 1.197 K_0$. The initial unstable
spheroid evolved in Fig 1. is indicated by the solid dot. For evolution with ``rapid cooling'' it evolves
vertically downward to a Jacobi ellipsoid of lower energy (solid triangle). For evolution with
``no cooling'' the star does not move in the figure, but it evolves to a triaxial ellipsoid on the equilibrium
branch with higher entropy, $K > K_0$. Although viscosity drives the instability, the final states are independent of its magnitude.
\label{fig2}}
\end{figure}

Performing numerous simulations of this type, we conclude that compressible 
Maclaurin spheroids are unstable to the {\it secular} bar-mode instability
whether or not they reside in the ``rapid-cooling'' or ``no-cooling''
regimes. Moreover, we find numerically that the onset of 
the instability coincides exactly with the bifurcation point along the 
compressible, isentropic Maclaurin curve, as in the case of incompressible spheroids. 
In the compressible ellipsoid approximation,
this point is located at 
$e=0.8127$ and $\mathcal{T}/|W| = 0.138$, independent of $n$ (LRS 1993).
However, the evolutionary tracks and final
Jacobi equilibrium configurations do depend on the cooling regime,
as described above. 


Finally, we recall that sequences of uniformly rotating polytropes
constructed from the exact (as opposed to the ellipsoidal)
equations of hydrostatic equilibrium do not
have a point of bifurcation when $n \geq 0.808$. We thus expect that
the results found here for compressible ellipsoids also apply to exact, uniformly rotating
polytropes, but only for configurations with $n \leq 0.808$. But they may also
apply to a wide class of differentially rotating polytropes, provided  $\mathcal{T}/|W| \gtrsim 0.14$.

\subsection{Determining the Final Equilibrium State}

Knowing that an unstable, compressible Maclaurin spheroid
evolves to a compressible ellipsoid allows us to
determine the final equilibrium state without
having to track the secular evolution. Once  
the appropriate cooling regime is assigned, the final state
is uniquely determined by the location of the initial spheroid 
along the Maclaurin equilibrium curve, as was illustrated in Fig. 2. 
Not only can we determine the size and shape of the final ellipsoid, and 
all other global parameters that characterize the final
equilibrium configuration, but we also can 
calculate its final thermodynamic state, i.e, its
entropy parameter $K$, independent of the magnitude of the viscosity.
It is the set of conserved quantities that allows us to determine
the final state uniquely, given the initial model.

\begin{deluxetable*}{cccccccccc}
\tablecaption{Endpoint Jacobi States for $n=1~^{\rm a}$}
\tablewidth{8.0in}
\tablehead{
\multicolumn{4}{c}{Initial Maclaurin Spheroid} & 
\colhead{} &
\multicolumn{5}{c}{Final Jacobi Ellipsoids} \\
\colhead{$e~^{\rm b}$} & \colhead{$\mathcal{T}/|W|$} & \colhead{$\bar J$} & 
\colhead{$\bar E$} & 
\colhead{} &
\colhead{$a_2/a_1$} & \colhead{$a_3/a_1$} &
\colhead{$R/R_0$} & \colhead{$\bar E$} & \colhead{$K/K_0$}}
\startdata
~~0.8127*     & 0.138         & 0.299        & -0.382       &  & 1.000        & 0.583        & 1.190  & -0.382 & 1.000  \\
              &               &              &              &  & 1.000        & 0.583        & 1.190  & -0.382 & 1.000  \\
0.8200        & 0.142         & 0.307        & -0.378       &  & 0.794        & 0.515        & 1.195  & -0.378 & 1.000  \\
              &               &              &              &  & 0.795        & 0.515        & 1.196  & -0.378 & 1.000  \\
0.8400        & 0.154         & 0.328        & -0.366       &  & 0.633        & 0.449        & 1.213  & -0.366 & 1.000  \\
              &               &              &              &  & 0.635        & 0.450        & 1.215  & -0.366 & 1.004  \\
0.8600        & 0.168         & 0.353        & -0.352       &  & 0.537        & 0.402        & 1.234  & -0.353 & 1.000  \\
              &               &              &              &  & 0.539        & 0.403        & 1.238  & -0.352 & 1.008  \\
0.8800        & 0.184         & 0.382        & -0.336       &  & 0.462        & 0.362        & 1.258  & -0.339 & 1.000  \\
              &               &              &              &  & 0.468        & 0.366        & 1.274  & -0.336 & 1.029  \\
0.9000        & 0.203         & 0.418        & -0.317       &  & 0.398        & 0.324        & 1.286  & -0.324 & 1.000  \\
              &               &              &              &  & 0.407        & 0.330        & 1.318  & -0.317 & 1.058  \\
0.9200        & 0.225         & 0.463        & -0.295       &  & 0.342        & 0.288        & 1.322  & -0.306 & 1.000  \\
              &               &              &              &  & 0.354        & 0.296        & 1.381  & -0.295 & 1.107  \\
0.9400        & 0.252         & 0.524        & -0.266       &  & 0.287        & 0.250        & 1.369  & -0.284 & 1.000  \\
              &               &              &              &  & 0.304        & 0.262        & 1.479  & -0.266 & 1.197  \\
~~~0.9529**   & 0.274         & 0.578        & -0.244       &  & 0.249        & 0.222        & 1.410  & -0.266 & 1.000  \\
              &               &              &              &  & 0.276        & 0.242        & 1.577  & -0.244 & 1.303  \\
0.9600        & 0.288         & 0.615        & -0.229       &  & 0.232        & 0.209        & 1.437  & -0.256 & 1.000  \\
              &               &              &              &  & 0.257        & 0.228        & 1.649  & -0.229 & 1.383  \\
0.9800        & 0.340         & 0.789        & -0.173       &  & 0.169        & 0.157        & 1.561  & -0.214 & 1.000  \\
              &               &              &              &  & 0.206        & 0.188        & 2.050  & -0.173 & 1.924  \\
0.9900        & 0.381         & 0.989        & -0.129       &  & 0.128        & 0.122        & 1.692  & -0.180 & 1.000  \\
              &               &              &              &  & 0.179        & 0.166        & 2.647  & -0.129 & 2.973  \\
0.9950        & 0.413         & 1.221        & -0.094       &  & 0.093        & 0.091        & 1.844  & -0.148 & 1.000  \\
              &               &              &              &  & 0.157        & 0.148        & 3.477  & -0.094 & 4.806  \\
0.9990        & 0.458         & 1.892        & -0.045       &  & 0.050        & 0.050        & 2.231  & -0.095 & 1.000  \\
              &               &              &              &  & 0.139        & 0.132        & 6.956  & -0.045 & 17.82~~  \\
1.~~~~~~      & 0.5~~~        & $\infty$     &  0.~~~~      &  & 0.~~~~       & 0.~~~~       & $\infty$ & 0.~~~~  & 1.~~~~  \\
              &               &              &              &  & 0.~~~~       & 0.~~~~       & $\infty$ & 0.~~~~  & $\infty$~~~ \\
\enddata

\tablenotetext{a}{$\bar J$, $\bar E$, and $R_0$ 
are defined in eqns.~(\ref{dim}) and ~(\ref{rad}); $R=(a_1 a_2 a_3)^{1/3}$}

\tablenotetext{b}{One asterisk marks the secular instability point, two the dynamical instabilty point.}
\end{deluxetable*}

First consider the determination of  the final state of an unstable
spheroid evolving in the ``rapid-cooling'' regime. The configuration
evolves at fixed $M$, $J$ and $K = K_0$. Thus the following 
dimensionless ratio is also constant:
\begin{equation} \label{final0}
\frac{1}{\kappa_n} \left( 1-n/5 \right) \left(\frac{J^2}{M^3 R_0} \right)
= {\rm constant} = \hat J^2 \hat R^{(n/(3-n))}.
\end{equation} 
Here $\kappa_n$ is a polytropic structure constant of order unity
which depends on the polytropic index $n$ (see LRS 1993 eqn. 3.14 and
Table 1).
The term on the right-hand side of eqn.~(\ref{final0}) is expressed
in terms of the universal dimensionless parameters $\hat J$ and
$\hat R$ each of which, like $\mathcal{T}/|W|$,  uniquely specifies a 
configuration along 
either a compressible Maclaurin {\it or} a Jacobi sequence, independent of $n$.
These parameters
were introduced by LRS (1993), (see their eqn (3.27)) where they are 
defined according to
\begin{eqnarray} \label{Jhat}
\hat J^2 &=& \frac{1}{\kappa_n} \left( 1-n/5 \right) 
\left(\frac{J^2}{M^3 R} \right), \\
\hat R &=& \left( \frac{R}{R_0} \right) ^{(3-n)/n},  \nonumber
\end{eqnarray}
where $R=(a_1 a_2 a_3)^{1/3}$. These universal parameters
are tabulated along a Maclaurin sequence in Table 2 and
along a Jacobi sequence in Table 4 of their paper. The significance
of the product on the right-hand side of eqn.~(\ref{final0}) is
that (1) it uniquely specifies a model along either a Maclaurin or
Jacobi sequence, and (2) once the initial Maclaurin model is specified,  
this product (unlike $\hat J$ or $\hat R$ separately) 
remains constant during ``rapid-cooling''  evolution and thereby
uniquely determines the final Jacobi model.

\begin{deluxetable*}{cccccccccc}
\tablecaption{Endpoint Jacobi States for $n=0.5~^{\rm a}$}
\tablewidth{6.0in}
\tablehead{
\multicolumn{4}{c}{Initial Maclaurin Spheroid} & 
\colhead{} &
\multicolumn{5}{c}{Final Jacobi Ellipsoids} \\
\colhead{$e~^{\rm b}$} & \colhead{$\mathcal{T}/|W|$} & \colhead{$\bar J$} & 
\colhead{$\bar E$} & 
\colhead{} &
\colhead{$a_2/a_1$} & \colhead{$a_3/a_1$} &
\colhead{$R/R_0$} & \colhead{$\bar E$} & \colhead{$K/K_0$}}
\startdata
 ~~0.8127*    & 0.138         & 0.299        & -0.450       &  & 1.000        & 0.583        & 1.072   & -0.450 & 1.000  \\
              &               &              &              &  & 1.000        & 0.583        & 1.072   & -0.450 & 1.000  \\
0.8200        & 0.142         & 0.306        & -0.446       &  & 0.796        & 0.516        & 1.074   & -0.446 & 1.000  \\
              &               &              &              &  & 0.796        & 0.516        & 1.074   & -0.446 & 1.000  \\
0.8400        & 0.154         & 0.325        & -0.435       &  & 0.637        & 0.451        & 1.080   & -0.436 & 1.000  \\
              &               &              &              &  & 0.638        & 0.451        & 1.082   & -0.435 & 1.007  \\
0.8600        & 0.168         & 0.347        & -0.423       &  & 0.541        & 0.405        & 1.087   & -0.423 & 1.000  \\
              &               &              &              &  & 0.542        & 0.405        & 1.088   & -0.423 & 1.003  \\
0.8800        & 0.184         & 0.373        & -0.408       &  & 0.469        & 0.366        & 1.095   & -0.412 & 1.000  \\
              &               &              &              &  & 0.474        & 0.369        & 1.107   & -0.408 & 1.057  \\
0.9000        & 0.203         & 0.403        & -0.391       &  & 0.406        & 0.329        & 1.105   & -0.397 & 1.000  \\
              &               &              &              &  & 0.415        & 0.334        & 1.128   & -0.391 & 1.115  \\
0.9200        & 0.225         & 0.441        & -0.370       &  & 0.350        & 0.293        & 1.116   & -0.381 & 1.000  \\
              &               &              &              &  & 0.362        & 0.302        & 1.158   & -0.370 & 1.217  \\
0.9400        & 0.252         & 0.489        & -0.344       &  & 0.297        & 0.257        & 1.130   & -0.361 & 1.000  \\
              &               &              &              &  & 0.315        & 0.269        & 1.206   & -0.344 & 1.411  \\
~~~0.9529**   & 0.274         & 0.531        & -0.322       &  & 0.264        & 0.233        & 1.142   & -0.346 & 1.000  \\
              &               &              &              &  & 0.285        & 0.248        & 1.254   & -0.322 & 1.647  \\
0.9600        & 0.288         & 0.559        & -0.308       &  & 0.244        & 0.218        & 1.150   & -0.336 & 1.000  \\
              &               &              &              &  & 0.268        & 0.236        & 1.290   & -0.308 & 1.849  \\
0.9800        & 0.340         & 0.680        & -0.253       &  & 0.185        & 0.171        & 1.183   & -0.299 & 1.000  \\
              &               &              &              &  & 0.220        & 0.200        & 1.482   & -0.253 & 3.377  \\
0.9900        & 0.381         & 0.805        & -0.206       &  & 0.144        & 0.137        & 1.215   & -0.267 & 1.000  \\ 
              &               &              &              &  & 0.192        & 0.176        & 1.742   & -0.206 & 7.058  \\
0.9950        & 0.413         & 0.935        & -0.167       &  & 0.118        & 0.113        & 1.246   & -0.241 & 1.000  \\ 
              &               &              &              &  & 0.176        & 0.163        & 2.083   & -0.167 & 16.45~~~  \\
0.9990        & 0.458         & 1.254        & -0.104       &  & 0.074        & 0.073        & 1.316   & -0.190 &  1.000 \\
              &               &              &              &  & 0.151        & 0.142        & 3.188   & -0.104 & 128.6~~~~~ \\
1.~~~~~~      & 0.5~~~        & $\infty$     &  0.~~~~      &  & 0.~~~~       & 0.~~~~       & $\infty$ & 0.~~~~  & 1.~~~~  \\
              &               &              &              &  & 0.~~~~       & 0.~~~~       & $\infty$ & 0.~~~~  & $\infty$~~~ \\
\enddata

\tablenotetext{a}{$\bar J$, $\bar E$, and $R_0$ are defined in eqns.~(\ref{dim}) and ~(\ref{rad}); $R=(a_1 a_2 a_3)^{1/3}$}

\tablenotetext{b}{One asterisk marks the secular instability point, two the dynamical instabilty point.}

\end{deluxetable*}

Next consider how the final state of an unstable
spheroid evolving in the ``no-cooling'' regime is determined. In this
case the configuration
evolves at fixed $M$, $J$ and $E$, while $K$ increases. Suppose we
guess the final value of $K$, setting $K = K^{\rm guess} > K_0$.
Then from eqn.~(\ref{rad}) this guess yields a new value for
$R_0$ given by
\begin{equation} \label{rguess}
R^{\rm guess}_0 = R_0 \left( \frac{K}{K_0} \right)^{n/(3-n)}.
\end{equation}
Proceeding as in the previous case, we use $R_0^{\rm guess}$ in 
eqn.~(\ref{final0}) to determine the product 
[$\hat J^2 \hat R^{(n/(3-n))}]_{\rm guess}$. This parameter specifes
a compressible ellipsoid. In general, however, the total energy
of the ellipsoid will not be equal to the energy of the initial,
unstable spheroid. Improved guesses of $K^{\rm guess}$ are required
to guarantee energy conservation. After a few iterations, the process
converges to the desired higher entropy, compressible 
ellipsoid with the same energy as the original spheroid.

We have applied the above prescription to determine in the compressible 
ellipsoidal approximation the endpoint
states of all unstable $n=1$ spheroids beyond the point of bifurcation.
Results are listed in Table 1. For each unstable spheroid, the top
row applies to the ``rapid-cooling'' regime, while the bottom row
applies to the ``no-cooling'' regime. We note that the two endpoint states 
derived for the unstable spheroid with
$e = 0.94$ are consistent with the results found by tracking the full
secular evolution of this model (slightly perturbed), as shown in Figs 1 and 2.
To assess the dependence on the polytropic index of the star, we determine
the endpoint states for unstable spheroids with a different polytropic index, 
$n=0.5$, in Table 2.
For evolution with ``rapid cooling'', stars with higher $n$ 
(i.e. softer equations of state and greater central concentration)
lose greater fractional energy than stars
with the same initial eccentricity (i.e. $\mathcal{T}/|W|$), but smaller $n$. For evolution with ``no cooling'', stars with 
higher $n$ experience less entropy increase (as measured by the 
increase in $K$) than stars
with smaller $n$. The reason for the entropy dependence is that, 
according to eqn.~(\ref{u}),  $U \propto k_1 K$ and
$k_1 \sim n$, so that a higher value of $K$ is necessary to heat a star
to the same $U$ for smaller $n$. Finally, we note that
the expansion of an unstable star evolving with ``no cooling'' 
in comparison with one evolving with ``rapid cooling'' (measured by the
ratio of their mean radii $R$) is larger for larger $n$, for the same initial
eccentricity.

\section{Point of Onset of the Bar-Mode Instability}

Our numerical integrations of the ellipsoidial dynamical equations with
viscosity show that the point of onset of the secular bar-mode 
instability in compressible spheroids 
coincides with the bifurcation point, where $\mathcal{T}/|W| = 0.1375$, 
independent of the polytropic index $n$ {\it and} 
the cooling regime. LRS (1993) provided a
rigorous proof of the first part of this assertion: they  
took the entropy parameter $K$ to be strictly constant, effectively assuming
``rapid cooling'',  and 
proved that the point of secular instability coincides with the 
bifurcation point for all $n$ (see Section 6.2 of LRS 1993).
We now sketch how this proof can be generalized to 
include the ``no-cooling'' regime as well.

The proof of LRS is based on an energy variational
principle and an energy functional for compressible ellipsoids.
The functional is of the form 
\begin{equation} \label{efunc}
E = E(\rho_c, \lambda_1, \lambda_2; M, J, K) = U + W + \mathcal{T},
\end{equation}
where $U$, $W$ and $\mathcal{T}$ are the internal, gravitational potential and
rotational kinetic energies of a compressible ellipsoid, not necessarily
in equilibrium. The ellipsoid of index $n$ is  
parametrized by the
central density $\rho_c$ and the two oblateness parameters
$\lambda_1 \equiv (a_3/a_1)^{2/3}$ and $\lambda_2 \equiv (a_3/a_2)^{2/3}$.
The equilibrium configuration of fixed $M$, $J$ and $K$ (i.e. fixed entropy) 
is determined by extremizing the energy according to
\begin{equation} \label{equil}
\frac{\partial E}{\partial \alpha_i} = 0, ~~~~i = 1 - 3,
\end{equation}
where $\{ \alpha_i \} = \{ \rho_c, \lambda_1, \lambda_2 \}$. 
Solving these equations simultaneously yields equilibrium relations
for compressible Jacobi ellipsoids. Stability requires that an
equilibrium configuration correspond to a true {\it minimum} of the
total energy, that is, that all eigenvalues of the matrix 
$\left ( \partial^2 E / \partial \alpha_i \partial \alpha_j \right)_{\rm eq}$ 
be positive. The onset of instability along any one-parameter sequence
of equilibrium configurations can be determined from the condition
\begin{equation} \label{det}
{\rm det} \left ( \frac{\partial ^2 E}{\partial \alpha_i \partial \alpha_j} \right )_{\rm eq}
= 0, ~~~~i,j = 1-3 ~.
\end{equation}
When this condition is first satisfied along the sequence, one of the
eigenvalues must change sign. By using the energy functional
for a compressible Jacobi ellipsoid given by eqn.~(\ref{efunc}), but
evaluating the determinant along the equilibrium Maclaurin sequence
(for which $\lambda_1 = \lambda_2$), LRS found that compressible
spheroids become secularly unstable to triaxial deformations at the
bifurcation point, where $\mathcal{T}/|W| = 0.1375$, independent of $n$.

Now suppose we allow for variations in which the entropy and entropy parameter $K$ is 
not fixed, but is allowed to increase, as is the case for evolution with ``no cooling''. 
In this situation, the relevant functional that must be minimized in a variational principle is 
the effective {\it free} energy of the configuration, 
\begin{equation} \label{free}
F = E - \langle T \rangle S,
\end{equation}
where the entropy, or equivalently, the entropy parameter $K$, must now be 
treated as another parameter 
that is allowed to vary:
$E = E(\rho_c, \lambda_1, \lambda_2, K; M, J)$. 
The parameter $K$ appears explicitly
in $E$ only through the internal energy $U$, given by eqn.~(\ref{u}). However,
eqn.~(\ref{kdot}) shows that the free energy $F$ does not change when $K$ changes,
while holding the other parameters fixed:
\begin{equation} \label{dF}
\frac{\partial F}{\partial K}  =  \frac{\partial E}{\partial K} - \langle T \rangle \frac{\partial S}{\partial K} 
            =  \frac {U}{K} - U \frac {\partial \ln K}{\partial K}  
            =  0 ~. 
\end{equation}
Similarly,
\begin{equation} \label{ddF}
\frac {\partial^2 F}{\partial \alpha_i \partial K} = 0 = 
\frac {\partial^2 F}{\partial K^2} = 0 , ~~~~i = 1-3 ~.
\end{equation} 
Hence the set of parameters $\{ \alpha_i \}$
found to extremize $E$ and identify its local minimum when holding $K$ fixed 
is the same set which extremizes $F$ and identifies its local minimum 
as $K$ is allowed to vary.  This set of parameters determines the onset of instability. 
Thus, compressible spheroids become secularly unstable 
to triaxial deformations at the
bifurcation point, where $\mathcal{T}/|W| = 0.1375$, independent of $n$
{\it and} independent of the degree of cooling.

\section{Summary and Future Work}

We have analyzed the secular bar instability driven by viscosity in rapidly rotating,
Newtonian stars. We have adopted the formalism of LRS, who model rotating stars as
compressible ellipsoids governed by a polytropic equation of state. However, we have extended 
their treatment to incorporate changes in the fluid entropy as the star evolves:
the gas is allowed to heat up by viscous dissipation and cool down by emission. We treated the
stars in two extreme opposite limits: one in which the cooling timescale is very much shorter than the
heating timescale (``rapid cooling'') and the other in which the timescale inequality is reversed (``no cooling''). 
Our numerical integrations of the ellipsoidal evolution equations show that,
in both limits, a uniformly rotating star spinning sufficiently rapidly is secularly unstable to
nonaxisymmetric perturbations to a bar. The point of onset of the bar instability along an 
equilibrium sequence of uniformly
rotating, compressible Maclaurin spheroids, parametrized by $\mathcal{T}/|W|$, occurs precisely at the bifurcation point 
($\mathcal{T}/|W| = 0.1375$)  in both regimes. 
This numerical finding is confirmed by a stability analysis based on 
minimizing the free energy functional of the star. However, the final equilibrium state of an unstable spheroid is different 
in the two regimes: for the ``rapid-cooling'' regime, the final configuration is a compressible Jacobi ellipsoid 
with the a lower energy but the same entropy as the initial unstable spheroid, while for the ``no-cooling'' regime, 
the final state is a Jacobi ellipsoid with the same energy but a higher entropy than the spheroid. The final states can
be calculated from the conserved quantities without having to track the evolution of the unstable stars.

As described in the Introduction and in LRS (1993,1994), 
rotating compressible ellipsoids mimic many of the same equilibrium and stability properties  
of rotating stellar models that obey the exact hydrodynamic equations. Not surprisingly, this concordance
is particularly close for stars that are governed by stiff equations of state (e.g. small polytropic
indicies $n$) and are not very centrally condensed, since ellipsoids provide exact solutions for homogeneous,
incompressible Newtonian stars.  It is therefore relevant
to examine the degree to which the new results found in this analysis and summarized above also apply to
exact compressible stellar models. We are preparing to study the secular bar-mode instability in 
uniformly and differentially rotating, compressible stars by performing numerical simulations  
that solve the exact Navier-Stokes equations with viscosity. We will treat the problem 
both in Newtonian gravity and in full general relativity. We will follow not only the onset but also the nonlinear 
growth of the instability and evolve unstable configurations until they arrive at a final equilibrium state. 
A preliminary numerical study that illustrates the numerical approach of tracking 
secular (viscous) evolution by performing full hydrodynamical simulations is given in Duez et al. (2004), where 
we followed the secular evolution and determined the final fate 
of rapidly differentially rotating, ``hypermassive'' neutron stars in general relativity, 
both in axisymmetry and in $3+1$ dimensions. Employing a hydrodynamical code to follow secular
evolution is necessary, either when the quasi-static configurations
traversed along the secular evolutionary track are difficult to construct by solving the hydrostatic
equilibrium equations (e.g., due to complicated internal flow patterns), or when the 
secular evolution ultimately sends the fluid into rapid hydrodynamical motion (e.g., catastrophic collapse). 
Our experience suggests that the technique, though resource intensive, is quite promising, even in general
relativity.

\acknowledgments

It is a pleasure to thank M. Duez, Y.T. Liu and B. Stephens for valuable
discussions.  This work was supported in part by NSF Grants 
PHY-0205155 and PHY-0345151 and NASA Grant NNG04GK54G at the
University of Illinois at Urbana-Champaign.

\end{document}